\documentclass[12pt]{article}

\begin{document}

\title{Specific Heat of YBa$_2$Cu$_3$O$_{7 - {\rm
\delta}}$ Single Crystals:  Implications for the Vortex Structure}
\author{Kathryn A. Moler and Aharon Kapitulnik \\
{\it {\small Departments of Physics and Applied Physics,
Stanford, CA 94305}}
\bigskip \\
David J. Baar$\rm^{a,b}$, Ruixing Liang$\rm ^{a,b}$, and Walter
N. Hardy$\rm ^{a}$ \\
{\it {\small $\rm ^{a}$Department of Physics, University of British
Columbia, }} \\
{\it {\small Vancouver, BC, Canada V6T 1Z1,~and  }} \\
{\it {\small
$\rm ^{b}$Quantum Innovations, Inc., P. O. Box 78512, Vancouver,
BC, Canada V6T 2E7}}}
\date{To appear in the Proceedings of the Stanford Conference on
Spectroscopies of Novel Superconductors}
\maketitle

\begin{abstract} The anisotropy of the magnetic field dependence
of the specific heat of YBa$_2$Cu$_3$O$_{7 - {\rm \delta}}$ can
be used to identify different low-energy excitations, which
include phonons, spin-$
\frac{1}{2}$ particles, and electronic contributions.  With a magnetic
field H applied perpendicular to the copper oxide planes, we find
that the specific heat includes a linear-T term proportional to
$\sqrt{H}$.  The nonlinear field dependence of the density of
states at the Fermi level suggests that there are quasiparticle
excitations throughout the entire vortex, not just in the vortex
core.  The
$\sqrt{H} T$ term agrees quantitatively with G. Volovik's
prediction for a superconductor with lines of nodes in the gap.
A similar, but much smaller, effect is predicted for fields
parallel to the planes, and sensitive measurements of the
in-plane anisotropic magnetic field dependence of the specific
heat could be used to map out the nodes.

\end{abstract}

\section{Introduction} Among recent experiments supporting the
possibility of lines of nodes in the gap function of some of the
high-T$_c$ materials
\cite{Dwave}, measurements on single-crystal
YBa$_2$Cu$_3$O$_{7 - {\rm
\delta}}$ (YBCO) show a linear temperature dependence of the
penetration depth
\cite{Hardy}, as predicted for superconductors with lines of
nodes in the clean limit \cite{Annett}.  The density of states
$N(E)$ in such superconductors rises linearly with energy at the
Fermi level in zero field, which should result in a quadratic
term $\alpha T^2$ in the specific heat
\cite{Prohammer}.  A material with a finite $N(E_F)$, such as a
normal metal or a disordered superconductor with lines of nodes
\cite{Hirschfeld}, has a linear term
$\gamma T$ in the zero-field specific heat, with  $\gamma \propto
N(E_F)$.

G. Volovik predicted that for a superconducting state with lines of
nodes, the dominant contribution to the magnetic field dependence of
the density of states at the Fermi level is
$N(E_F,H)=\kappa N_n \sqrt{H/H_{c2}}$, where $N_n$ is the
normal-state DOS and $\kappa$ is a factor of order 1
\cite{Volovik}.  The factor $\kappa$ is defined by the vortex
lattice  structure, by the slope of the gap near the gap node, and
by possible structure in $N_n$.   The quasiparticles which
contribute to $N(E_F,H)$ are outside the vortex cores, and are
close to the nodes in momentum space.  By contrast, it is usually
thought that in a clean s-wave superconductor the quasiparticles
are confined to the vortex cores
\cite{Caroli}.

In previous papers \cite{MolerPRL,MolerMiami} we reported
measurements of the magnetic field dependence of the specific
heat of twinned and untwinned single crystals of
YBa$_2$Cu$_3$O$_{6.95}$ which were prepared identically to those
showing linear temperature dependence of the penetration depth
\cite{Hardy,Liang}.  These crystals have a linear term $\gamma
(0)T$ in the zero-field specific heat, a quadratic term $\alpha
T^2$ which is only a few percent of the total zero-field specific
heat, and an additional field-dependent linear-T term
$\gamma _{\perp} (H)T$ which obeys $\gamma _{\perp} (H)T=A
\sqrt{H}T$.  Two of these terms,
$\alpha T^2$ and $A \sqrt{H}T$, agree quantitatively with the
predictions of lines of nodes.

In this paper we summarize the previous analysis, including
alternative fits, and present  results on the sample dependence
of the zero-field linear term,
$\gamma (0)T$.  We also discuss the physical origin of the
$\sqrt{H}$ dependence of the density of states and the prospects
for observing this effect in a magnetic field applied parallel to
the plane.  In principle, measurements of the in-plane
anisotropic magnetic field dependence of the specific heat could
be used to map out the nodes.

\section{Technique} The specific heat was measured using a
relaxation method described in detail elsewhere \cite{Urbach}.
The sample was mounted on a sapphire substrate with a weak
thermal link (thermal conductance $\kappa _w$) to a
constant-temperature copper block.  At each temperature and
field, $\kappa _w$ was measured by applying power
$P=\kappa _w (T_{sample}-T_{block})$.  A smaller temperature
difference was then used to make a relaxation measurement of the
thermal time constant $\tau = C \kappa _w$, where $C$ is the
total heat capacity.  The precision of the measurement in the
temperature range 2--10 K is $\sim 0.5 \, \% $.  The addenda heat
capacity can be as high as half of the total heat capacity,
limiting the absolute accuracy in some measurements to $\sim 10
\, \% $.  The addenda specific heat is mostly due to phonons, so
that most of the systematic uncertainty effects only the Debye
$\beta T^3$ term.  Measurements of an empty substrate showed that
the addenda heat capacity does not depend on the applied magnetic
field within the precision of the data. The accuracy of the field
dependence of the specific heat is better than $1 \, \% $.

\section{Crystals} All of the samples are single crystals of
YBa$_2$Cu$_3$O$_{7 - {\rm \delta}}$ grown by the same flux-growth
technique \cite{Liang} as crystals which show linear temperature
dependence of the penetration depth \cite{Hardy}.  Results have
been obtained on two nominally pure, twinned single crystals (samples
T1 and T2).  A 0.8 mg piece was cleaved from one of the twinned
crystals, sample T1, and detwinned to create sample U1.  About
$20 \, \%$ of the surface of sample U1, near two of the corners,
had a high density of visible twins remaining after the
detwinning process. Sample T2 was mostly detwinned to create sample
U2, which had visible twin boundaries remaining.  Both U1
and U2 were reoxygenated subsequent to detwinning.

\section{Results} Figure \ref{fig:totalcp}
shows the specific heat of sample T1 in an 8 Tesla field applied
either parallel or perpendicular to the copper oxide planes.  The
total specific heat includes the Debye phonon contribution,
$\beta T^3$, which has a similar value in all three data sets;
the zero-field linear term
$\gamma (0)T$, which appears as a finite intercept on the
c/T-axis; a magnetic contribution from spin-$\frac{1}{2}$ particles,
$n c_{Schottky}(T,H)$, which appears as a low-temperature bump
peaking around 3-4 Kelvin in both parallel and perpendicular
fields; and a substantially increased linear term in
perpendicular field, $\gamma _{\perp} (H)T$.

\subsection{Perpendicular Field---Global Fit}
To model the perpendicular field dependence, identical analyses
are performed on data from samples T1 (H = 0, 0.5, 2, 4, 6, and 8
Tesla) and U1 (H = 0, 0.5, 2, 4, 6, 8, and 10 Tesla).  A global
fit is used to constrain the concentration of spin-$\frac{1}{2}$
particles, $n$, and the phonon specific heat, $\beta T^3$, to be
field-independent.  Allowing for the possibility of a small
quadratic term, $\alpha T^2$, as predicted for lines of nodes, the
zero-field specific heat is described by:
\begin{equation}  c(T,0) = \gamma (0)T+ \alpha T^2 + \beta T^3.
\label{eq:zero}
\end{equation} The specific heat for $H \geq 0.5$ T is described
by:
\begin{equation}  c(T,H) = \left( \gamma (0) + \gamma _{\perp} (H)
\right) T + \beta T^3 +nc_{Schottky}(g \mu _B H/k_BT) ,
\label{eq:h}
\end{equation} where
\begin{equation}  c_{Schottky}(x) = x^2 \frac{e^x}{(1+e^x)^2}.
\label{eq:schottky}
\end{equation}

All data sets are fit simultaneously to Equations
\ref{eq:zero}, \ref{eq:h}, and \ref{eq:schottky} with a Land\'{e}
g-factor of
$g=2$~\cite{MolerMiami}.  Thus, $\beta$ and $n$ are constrained
by multiple data sets, $\gamma (0)$ and
$\alpha$ are constrained by the zero-field data, and $\gamma
_{\perp} (H)T$ is allowed a different value at each field.  The
statistical confidence levels are determined by a bootstrapping
technique
\cite{Efron}, in which alternative data sets are randomly chosen
from the original data set and the fitting procedure is
repeated.  This procedure for determining the statistical error
bars avoids making any assumption about the statistical error
distribution of the data set.

The parameters determined from this fit are shown in
Table~\ref{table:global}, with $90 \, \% $ statistical confidence
levels. The coefficient of the Schottky anomaly, $n$, indicates
$0.1 \, \%$ spin-$\frac{1}{2}$ particles per Copper.  The coefficient
of the Debye term,
$\beta = 0.39 \,\rm mJ/molK^4$, corresponds to a Debye
temperature of 400 K. The difference in the Debye term between
the two samples is within the systematic error of the Debye term
of the addenda.  The rms deviation of the data from the fit is
$0.8 \,
\% $ for T1 and $2.7 \, \% $ for U1.  The higher scatter in the
data for the untwinned crystal U1 results from the smaller sample
size.  The data and fits for both samples are shown in Figure
\ref{fig:fits}, with the phonon specific heat ($\beta T^3$)
subtracted.

\begin{table}[h]
\caption{Global Fit Parameters (mJ, mol, K, and T)}
\begin{center}
\begin{tabular}{c|cc}
  & Sample T1  & Sample U1 \\ \hline \hline
$\gamma (0)$ &	$3.0 \pm 0.1$	& $2.1+0.1-0.2$ \\
$n$	& $24 \pm 1$ &	$ 23 \pm 1 $ \\
$\beta$	& $0.392 \pm 0.001$ & $0.380 \pm 0.004$ \\
$\alpha$	& $0.11 \pm 0.02$	& $0.10 \pm 0.06$ \\
$\gamma _{\perp} (H)$ &	$0.91 \sqrt{H}$ &	$0.88 \sqrt{H}$ \\
\end{tabular}
\end{center}
\label{table:global}
\end{table}

In the measurements on single crystals of YBCO reported here, the
phonon specific heat obeys the Debye $\beta T^3$ law up to 8 K.
In the zero-field data set on sample T1, for example, fits from 2
to 4 K and from 4 to 7 K give the same value of $\beta$ within
statistical error, indicating that higher powers of temperature
are not necessary to describe the data.  Above 8 K, the
deviations from
$\beta T^3$ are better described by a gapped excitation such as an
optical phonon mode than by a sum of $T^5$, $T^7$, and
$T^9$ terms.  In contrast, the data on many polycrystalline
samples require a large
$T^5$ term to describe the phonon specific heat even below 5
K~\cite{Reviews}.  In order to avoid the extra parameters
necessary to describe the phonon specific heat above 8 K, all of
the fits reported in this paper are restricted to $T \leq 7$~K.

The field-dependent linear term, $\gamma _{\perp} (H)T$, has a
nonlinear dependence on field (Figure \ref{fig:gammah}).  This
nonlinear field dependence is well-described by $\gamma_{\perp}
(H) = A
\sqrt{H}$, with $A = 0.9$ mJ/molK$^2$T$^{1/2}$ for both samples
T1 and U1.

\subsection{Perpendicular Field---Alternative Fits}

The qualitative nonlinearity of $\gamma_{\perp}(H)$ is robust to
the assumptions used to describe the total specific heat, and is
well described by $\gamma_{\perp}(H) =A \sqrt{H}$.  To check the assumptions of
the global
fit and the interdependence of the
$\alpha T^2$ and $A \sqrt{H}T$ terms, each data
set of sample T1 is fit independently to Equation~\ref{eq:h}.  There
are thus three parameters each for five data sets ($H=0.5$--8~T)
and two parameters for the zero-field data set (note that
$c_{Schottky}(T,0)=0$), totaling seventeen parameters for six
data sets. No zero-field quadratic term
was allowed in this fit.  A small quadratic term, if present,
could easily be absorbed into a slightly larger cubic term, and
this fit results in a $\beta (H=0)$ which is $\sim 5 \,
\%$ higher than $\beta (H \neq 0)$.  The
coefficients of the linear term are shown in
Figure~\ref{fig:gammaalt}a with error bars at the statistical
$90 \, \%$ confidence level.  Fitting the coefficients of the
linear term to
$\gamma_{\perp} (H) = A \sqrt{H}T$ gives $A = 0.89$
mJ/molK$^2$T$^{1/2}$.

In another check which avoids the zero-field data entirely, the
five data sets with $H
\geq 0.5$ T are globally fit to Equation~\ref{eq:h}, again keeping
$\beta$ and $n$ independent of field.  There are thus two global
parameters, plus values of
$(\gamma (0) + \gamma_{\perp} (H))$ in five fields
(Figure~\ref{fig:gammaalt}b).  The resulting linear-T term is
better described by a $\sqrt{H}$ dependence than by an $H$
dependence, and from the fit shown in
Figure~\ref{fig:gammaalt}b,  $A = 0.87$ mJ/molK$^2$T$^{1/2}$.
This fit also returns an extrapolated value of
$\gamma (0) = 3.1$ mJ/molK$^2$, which is in agreement with the
value determined by independently fitting just the zero-field
data.

\subsection{Parallel Field}
The data on sample T1 in an 8
Tesla parallel field show a linear-T term, $\gamma
_{\parallel}(H)T$, which is increased by about 0.5 mJ/molK$^2$ from
$\gamma (0)$, as determined by fitting the data in Figure
\ref{fig:totalcp} to Equation \ref{eq:h}.  Because of uncertainties
in the Schottky anomaly and the sample alignment, the results are
also consistent with no increase in the parallel-field linear term,
$\gamma _{\parallel}(H)T$.

\subsection{Zero-field Linear Term} The coefficient of the
zero-field linear term, $\gamma (0)T$, in the specific heat of
each of the four samples is shown in Table~\ref{table:samples}.
The zero-field linear term for the two twinned samples, samples
T1 and T2, is $\gamma (0) = 3$  mJ/molK$^2$, while in both
untwinned samples, U1 and U2, the linear term is reduced to
$\gamma (0) = 2$  mJ/molK$^2$.  In contrast, most YBCO samples
have
$\gamma (0)
\geq 4$  mJ/molK$^2$~\cite{Reviews}. The coefficient of the
linear term in the normal state of optimally-doped YBCO, $\gamma
_n$, has been shown by other measurements to be about 20
mJ/molK$^2$~\cite{Reviews,Loram}.  The residual density of states
$N(E_F)$ in the superconducting state is therefore about $15 \,
\% $ of the normal-state density of states for the twinned
samples and about $10 \, \% $ for the untwinned samples.

\begin{table}[h]
\caption{Zero-Field Linear Term, $\gamma (0)$ (mJ/molK$^2$)}
\begin{center}
\begin{tabular}{cccccc}  Sample  & ${7 - {\rm
\delta}}$ & Untwinned & $\gamma (0)$ &
$N(E_F)/N_n$ \\ \hline \hline T1    & 6.95 &  no  &
3.1 &
$\sim 15 \, \% $ \\  U1  & 6.95 &  yes  & 2.1 & $\sim
10 \, \%
$ \\  "  & 6.97 &  yes  & 1.9 & $\sim 10 \, \% $ \\
T2  & 6.95 &  no  & 2.8 & $\sim 15 \, \% $ \\  U2  & 6.95 &  yes  &
2 & $\sim 10 \, \% $ \\
\end{tabular}
\end{center}
\label{table:samples}
\end{table}

\section{Discussion}
\subsection{Comparison with Lines of Nodes}
The quadratic term is predicted to disappear in a magnetic field,
where the energy dependence of the density of states close to the
nodes,
$N(E,H=0) \propto |E-E_F|$, is replaced by a finite $N(E_F,H)$.
The $\alpha T^2$ term is comparable in magnitude to the systematic
uncertainty in the addenda $T^3$ phonon specific heat, but is
several times larger than the relatively small systematic
uncertainty of the field dependence, $< 1 \, \%$.  The global fit
is thus the only possible way to identify the quadratic term in
these data sets.  The slight positive slope of the zero-field data
set in Figure~\ref{fig:fits} can be described either as a phonon
term which is about $5 \, \%$ larger in zero field, which is
difficult to explain, or as  an $\alpha T^2$ term.

{}From the slope of the density of
states in a superconductor with lines of nodes it is possible to
predict $\alpha \approx \gamma_n/T_c \approx 0.2 \,
\rm mJ/molK^3$ for YBCO within factors of order
unity~\cite{MolerPRL}, in good agreement with the value $\alpha =
0.1 \,
\rm mJ/molK^3$ obtained from the global fits
(Table~\ref{table:global}) on samples U1 and T1.

For the field dependence of the density of
states, Volovik predicts $N(E_F,H)=\kappa N_n
\sqrt{H/H_{c2}}$~\cite{Volovik}, where $\kappa$ is of order
1 and is defined by the vortex lattice  structure, by the slope of
the gap near the gap node, and by possible structure in $N_n$.
Taking  $\gamma_n= 20 \,
\rm mJ/molK^2$~\cite{Reviews,Loram} and $H_{c2,\perp}=150 \, \rm T$
gives the prediction $\gamma_{\perp}(H)=A\sqrt{H}$
with $A=\kappa \ast 1.6 \,  \rm mJ/molK^2T^{1/2}$, in good
agreement with the value $A=0.9 \,  \rm
mJ/molK^2T^{1/2}$ found for both samples U1 and T1.

Both of the above comparisons used the measured value of
$\gamma_n$, the angular averaged normal state density of states.
Because both the $\alpha T^2$ and $A \sqrt{H} T$ terms are a
consequence of the nodes, these terms should be sensitive to
the normal-state density of states close to the nodes.  The
above comparisons may indicate that the density of states close to
the nodes is smaller than the angular average of the density of
states, as indicated by photoemission measurements
on BSCCO \cite{Dessau}.

These predictions do not distinguish between different types
of lines of nodes, such as d-wave or extended s-wave
\cite{Varma}.  In principle, this interpretation could also apply
to a gap function with no nodes, but with a very small minimum
gap, $\Delta \ll k_BT \approx 0.5 \, \rm meV$.  In order to
produce these results, such a gap function would need to have an
unusual energy dependence of the density of
states similar to that associated with lines of nodes, $N(E,H=0)
\propto |E-E_F|$.

A quadratic $T^2$ term has been known for some time in the
specific heat of heavy-fermion superconductors, and has recently
been reported in the specific heat of La$_{2-x}$Sr$_x$CuO$_4$
\cite{Momono}.  Two other recent works support the existence of a
$\sqrt{H}$ term in superconductors with lines of nodes: a
reanalysis of existing data on polycrystalline YBCO samples shows
a nonlinear $\gamma (H)$T term which appears consistent with
$\sqrt{H}$ \cite{Fisher}, with similar but less dramatic
curvature on Ca- and Sr-doped La$_2$CuO$_4$ samples, and a
$\sqrt{H}$ term has also been recently reported in the
low-temperature specific heat of UPt$_3$
\cite{Ramirez}.

\subsection{Residual Density of States}
The total residual density of states
$N(E_F)$ in the superconducting state, as determined from the
$\gamma(0) T$ term in the specific heat, is about
$15 \, \% $ of the normal-state density of states for the twinned
samples and about $10 \, \% $ for the untwinned samples
(Table~\ref{table:samples}). The residual density of states can also
be extracted from fitting the temperature dependence of the
penetration depth \cite{Hardy} to the d-wave model with scattering
of Hirschfeld {\em et al.\/}
\cite{Hirschfeld}.  In samples similar to the samples measured
here, these fits are inconsistent with a density of states which
is greater than $1 \, \% $ of $N_n$~\cite{Hirschfeld}.
It appears difficult to reconcile the zero-field
linear term $\gamma (0)T$ with the $\alpha T^2$ and $A
\sqrt{H}T$ terms, and with the interpretation of the temperature
dependence of the penetration depth.  It is possible that the
zero-field linear term has a separate origin from the excitations
which give rise to the quadratic term and the field dependence of
the specific heat.  This assumption leads to a self-consistent
analysis of the specific heat, and is given additional support by
the measurements on the untwinned crystal U1, in which $\gamma
(0)$ is decreased substantially but the $\alpha T^2$ and $A
\sqrt{H}T$ terms are unchanged within statistical error bars.

\subsection{Interpretation of the $A \protect\sqrt{H} T$ term}
Conventional high-$\kappa$ superconductors are expected to have a
linear-T term in the specific heat which is proportional to the
volume of normal material in the cores, $c
\approx \gamma_nTH/H_{c2}$~\cite{Caroli,Fetter}, and such an $HT$
term has been observed in A15 superconductors~\cite{Stewart}.
The analysis which predicts this
$HT$ term does not apply to cuprate superconductors for two
reasons.  First, the small coherence length of the cuprates may
result in a large excitation gap for quasiparticles which are
confined to the vortex cores
\cite{Caroli,Karrai}.  Secondly, Caroli, deGennes, and Matricon
\cite{Caroli} assumed a fully gapped superconductor in showing
that the quasiparticle excitations are confined to the vortex
core.

Some form of a linear-H term might be expected in the specific
heat for any excitations which are confined to the vortex cores.
Although there may in principle be some dependence of the core
excitations on magnetic field, it is likely that for
$H \ll H_{c2}$ the core excitations would be roughly independent
of field.  Thus, any signal resulting from these excitations
would simply be proportional to the number of cores, or linear in
H.  The nonlinear field dependence of the observed
$N(E_F,H)$ suggests excitations outside the vortex core.

The physical origin of the density of states predicted by Volovik
\cite{Volovik} is the same as the physical origin of the depairing
current in a conventional superconductor.  In the presence of a
supercurrent with velocity $\vec{v_s}$, the quasiparticle
excitation spectrum $\xi(\vec{k})$ is shifted by an amount
$\vec{k}
\cdot \vec{v_s}$.  Far from the vortex core in a fully gapped
superconductor, this shift is not large enough to change the
density of states at the Fermi level, which is given by
\begin{equation}  N(E_F)=\int \frac{d^3k}{(2 \pi)^3} \int d^2r
\delta
\left( \xi(\vec{k},\vec{r}) -\vec{k}
\cdot \vec{v_s} \right).
\label{eq:dos}
\end{equation} For a superconductor with lines of nodes, the
shift is significant everywhere that the superfluid velocity is
not zero.  Assuming a vortex superfluid flow
$v_s \propto 1/r$, the local density of states at the Fermi level
also falls off as $1/r$ (Figure \ref{fig:vortex}).  Integrating over
an entire vortex, with the intervortex spacing $R(H)$ as the
upper cutoff in the integral, gives a density of states per
vortex $N(E_F,H) \propto R(H)
\propto 1/\sqrt{H}$.  Multiplying by the number of vortices,
which is proportional to
$H$, then gives the total density of states $N(E_F,H) \propto
\sqrt{H}$.

The shift in the excitation spectrum depends on the angle between
the local superfluid velocity and the nodes, giving the vortex
the same symmetry as the gap (Figure \ref{fig:vortex}).  Because
the nodes are $90 ^{\circ}$ apart, the superfluid velocity can
never be perpendicular to all nodes, and there are no places
where $N(E_F)$ becomes zero.  Both the density of states, and in
a clean system the anisotropy in the density of states, extend
throughout the entire vortex.

B. Parks {\em et al.\/} \cite{Parks} have pointed out that their
measurements of vortex dynamics in YBCO thin films are best
explained by the presence of the extra quasiparticles expected
for a vortex with an anisotropic gap.  Recent theoretical work
suggests that the structure of the vortex core in a d-wave
superconductor is both interesting and
complicated~\cite{Volovik,Soininen,Berlinsky}.  We have
deliberately omitted the core region from
Figure~\ref{fig:vortex}, since we are not aware of any model for
the core which suggests that our measurements are probing the
core states.

\subsection{Predicted Parallel Field Dependence}

Because $N(E_F,H)$ depends on the orientation of the current
with respect to the nodes, improved measurements of the full
angular dependence $\gamma _{\theta,\varphi}(H)T$ should be
sensitive to the positions of the nodes. Define the
orientation of the applied field $\vec{H}$  as
$(\theta,\varphi)$, where $\theta$ is the polar angle measured
with respect to the c-axis and $\varphi$ is the azimuthal angle
measured with respect to the (110) crystalline axes.  For
tetragonal $d_{x^2-y^2}$ symmetry with antinodes along the Cu-O
bonds,
$\varphi=0$ corresponds to a field parallel to a node (Figure
\ref{fig:nodes}).  In this orientation,
$(\theta=\pi/2,\varphi=0)$, the currents in the ab-plane are
flowing parallel to a single node.  The density of states at the
Fermi level is
$\sqrt{2}$ larger at $(\theta=\pi/2,\varphi=\pi/4)$, where the
currents flow parallel to the antinodes (Figure
\ref{fig:nodes}):  in this orientation, the density of states
picks up contributions from all of the nodes.  The complete
in-plane angular magnetic field dependence of the density of
states is given by

\begin{equation}
N(0,H,\theta=\pi/2,\varphi)=\frac{\kappa \prime}{2}\left(|
\sin \varphi | + | \cos \varphi |
\right)N_F
\sqrt{\frac{H}{H_{c2\parallel}}},
\label{eq:nphi}
\end{equation}

as shown by Volovik \cite{VolovikUn}.  Using $\gamma_n = 20 \,
\rm mJ/molK^2$ and taking ${H}/{H_{c2
\parallel}} = 0.005$ as an experimentally accessible field, the
field-dependent linear term $\gamma _{\parallel}(H)$ will vary from
0.7 to 1.0 mJ/molK$^2$ as a function of the azimuthal angle
$\varphi$.  This predicted angular
dependence of the linear term in the specific heat may be within
achievable experimental resolution, although preliminary
measurements at 8 Tesla on a heavily twinned sample with the field
parallel to the twins and at a $45 \deg$ angle to the twins
did not detect any angular variation in $\gamma _{\parallel}(H)$.

\section{Summary} The residual density of states at the Fermi
level in the superconducting state of single-crystal
YBa$_2$Cu$_3$O$_{7 - {\rm
\delta}}$,
$N(E_F,H=0)/N_n$, is determined by the zero-field linear term
$\gamma(H=0)T$ in the specific heat.  $N(E_F,H=0)/N_n$ is
qualitatively larger than would be expected from fitting the
temperature dependence of the penetration depth~\cite{Hardy} to
expressions for lines of nodes with
scattering~\cite{Hirschfeld}.  In two twinned single crystals,
$\gamma(H=0)/\gamma_n
\approx 0.15$, while in two untwinned single crystals,
$\gamma(H=0)/\gamma_n
\approx 0.10$.

The specific heat also includes a $\gamma_{\perp}(H)T$ term,
which obeys
$\gamma_{\perp}(H) \approx \gamma_n \sqrt{H/H_{c2}}$ as predicted
for superconductors with lines of nodes in the gap
function~\cite{Volovik}.  This nonlinear field dependence of the
density of states suggests quasiparticle excitations outside the
vortex core, and appears to be independent of twinning, unlike
the zero-field linear term.  The in-plane angular magnetic field
dependence of the specific heat, which is smaller than the
perpendicular magnetic field dependence, is  predicted to be
sensitive to the locations of the nodes.

We thank M. R. Beasley, C. M. Varma, and especially G.
Volovik for many useful discussions, and R. A. Fisher, N. E. Phillips,
and J. Gordon for sharing their analysis prior to publication.

\pagebreak

\begin{figure}
\caption{Total specific heat at 0 and 8 Tesla for sample T1}
\label{fig:totalcp}
\end{figure}

\begin{figure}
\caption{Nonphonon specific heat for samples T1 and
U1, with the global fit described in the text.}
\label{fig:fits}
\end{figure}

\begin{figure}
\caption{Coefficient of the linear term for crystals T1
and U1, determined by the global fit described in the text.}
\label{fig:gammah}
\end{figure}

\begin{figure}
\caption{Coefficient of the linear term for crystal T1,
determined by a) independent fits in which all parameters might be
field-dependent, and b) a global fit with the
$H=0$ data set excluded}
\label{fig:gammaalt}
\end{figure}

\begin{figure}
\caption{Schematic of the local density of states $N(E_F)/N_n$
throughout a d-wave vortex.}
\label{fig:vortex}
\end{figure}

\begin{figure}
\caption{The azimuthal field angle $\varphi$ relative to a
tetragonal d-wave order parameter, and sketch of the current flows
associated with an in-plane vortex for $\varphi=0$ and
$\varphi=\pi/4$.}
\label{fig:nodes}
\end{figure}


\begin{thebibliography}{99}

\bibitem{Dwave} see e.g. D. Scalapino, {\em Physics Reports\/} {\bf
250}, (1995) 329; D. Pines and P. Monthoux,
{\em J. Phys. Chem. Solids\/},
to be published, 1995.

\bibitem{Hardy} W. N. Hardy {\em et al.\/},  {\em Phys. Rev.
Lett\/} {\bf 70}, (1993) 3999.

\bibitem{Annett} J. Annett {\em et al.\/},  {\em Phys. Rev. B\/}
{\bf 43}, (1991) 2778.

\bibitem{Prohammer} M. Prohammer {\em et al.\/},  {\em Phys. Rev.
B\/} {\bf 47}, (1993) 15152.

\bibitem{Hirschfeld} P. J. Hirschfeld. {\em et al.\/},  {\em
Phys. Rev. B\/} {\bf 50}, (1994) 10250.

\bibitem{Volovik} G. E. Volovik,  {\em JETP Lett.\/} {\bf 58},
(1993) 469.

\bibitem{Caroli} Caroli, deGennes, and Matricon,  {\em Phys.
Lett.\/} {\bf 9}, (1964) 307.

\bibitem{MolerPRL} K. A. Moler {\em et al.\/},  {\em Phys. Rev.
Lett.\/} {\bf 73}, (1994) 2744.

\bibitem{MolerMiami} K. A. Moler {\em et al.\/},  Proc. U-Miami
Workshop, to be published in {\em J. Supercond.\/}, (1995).

\bibitem{Liang} Ruixing Liang {\em et al.\/},  {\em Physica C\/}
{\bf 195}, (1992) 51.

\bibitem{Urbach} J. S. Urbach {\em et al.\/},  {\em Phys. Rev.
B\/} {\bf 39}, (1989) 12391.

\bibitem{Efron}  B. Efron,  {\em An introduction to the
bootstrap\/} (Chapman \& Hall, New York, 1993).

\bibitem{Reviews} see e.g. N. E. Phillips {\em et al.\/}, Prog.
in Low Temp. Phys. XIII,
 D. F. Brewer, ed., (Elsevier Science Publishers B. V., 1992); A.
Junod, Physical Properties of HTSC II, D. Ginsberg, ed.,  (World
Scientific,Singapore, 1990).

\bibitem{Loram} J. W. Loram {\em et al.\/}, {\em Phys. Rev.
Lett.\/} {\bf 71}, (1993) 1740.

\bibitem{Dessau} D. S. Dessau {\em et al.\/}, {\em Phys. Rev.
Lett.\/} {\bf 71}, (1993) 2781.

\bibitem{Varma} C. M. Varma, private communication; P. B.
Littlewood {\em et al.\/},  {\em Phys. Rev. Lett.\/} {\bf 63},
(1989) 26002.

\bibitem{Momono} N. Momono {\em et al.\/},  {\em Physica C\/}
{\bf 233}, (1994) 395.

\bibitem{Fisher}R. A. Fisher {\em et al.\/}, unpublished.

\bibitem{Ramirez} A. Ramirez {\em et al.\/}, {\em Phys. Rev.
Lett.\/} {\bf 74}, (1995) 1218.

\bibitem{Fetter} A. L. Fetter and P. Hohenberg, in:  Parks, ed.,
{\em Superconductivity \/} {\bf II} (1969).

\bibitem{Stewart} G. R. Stewart and B. L. Brandt, {\em Phys. Rev.
B\/} {\bf 29}, (1984) 3908.

\bibitem{Karrai}  K. Karrai {\em et al.\/}, {\em Phys. Rev.
Lett.\/} {\bf 69}, (1992) 152.

\bibitem{Parks} B. Parks {\em et al.\/},  {\em Phys. Rev.
Lett.\/}, to be published.

\bibitem{Soininen} P. I. Soininen {\em et al.\/},  {\em Phys.
Rev. B\/} {\bf 50}, (1994) 13883.

\bibitem{Berlinsky} A. J. Berlinsky {\em et al.\/},  preprint.

\bibitem{VolovikUn} G. E. Volovik, unpublished.

\end{thebibliography}
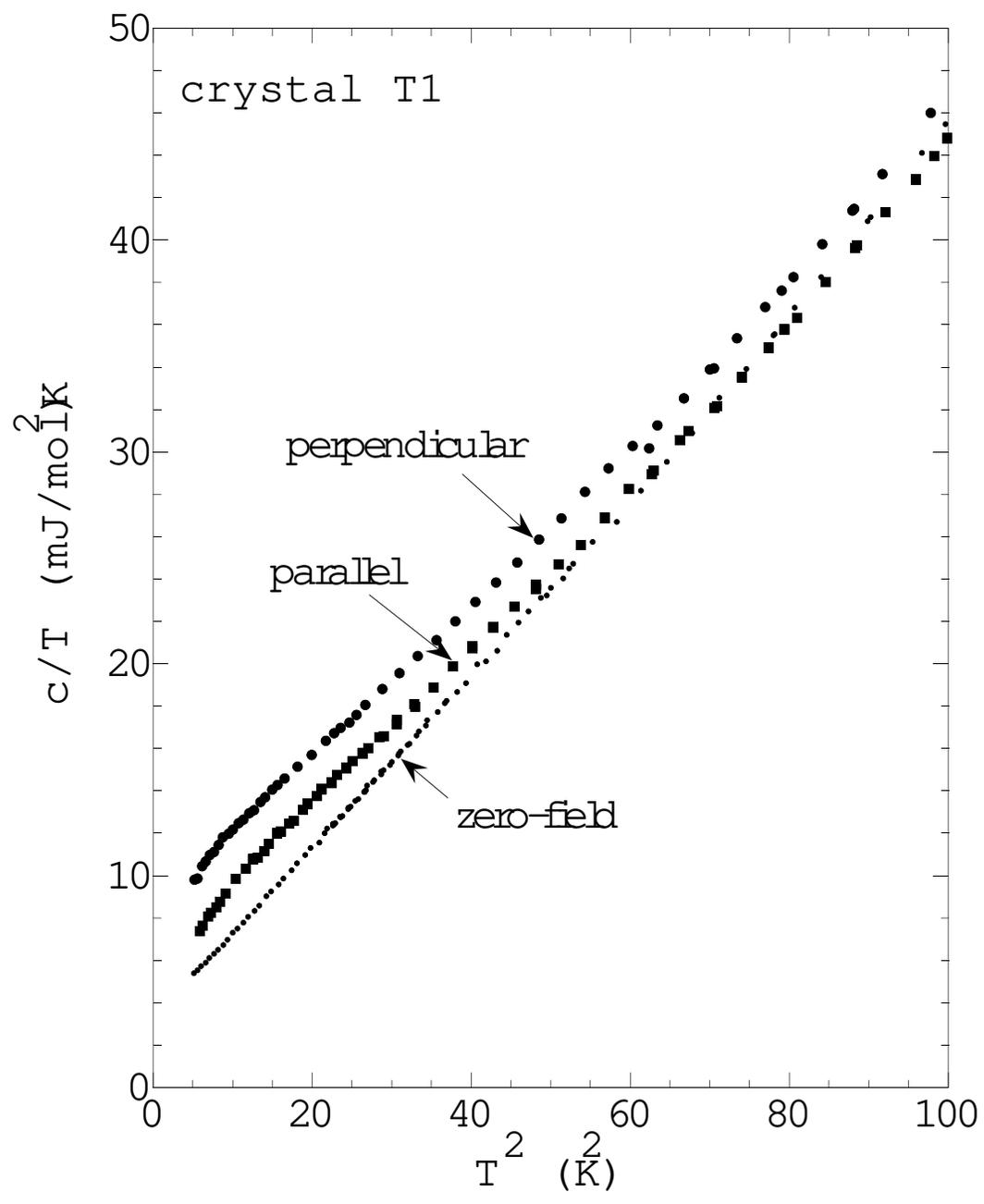
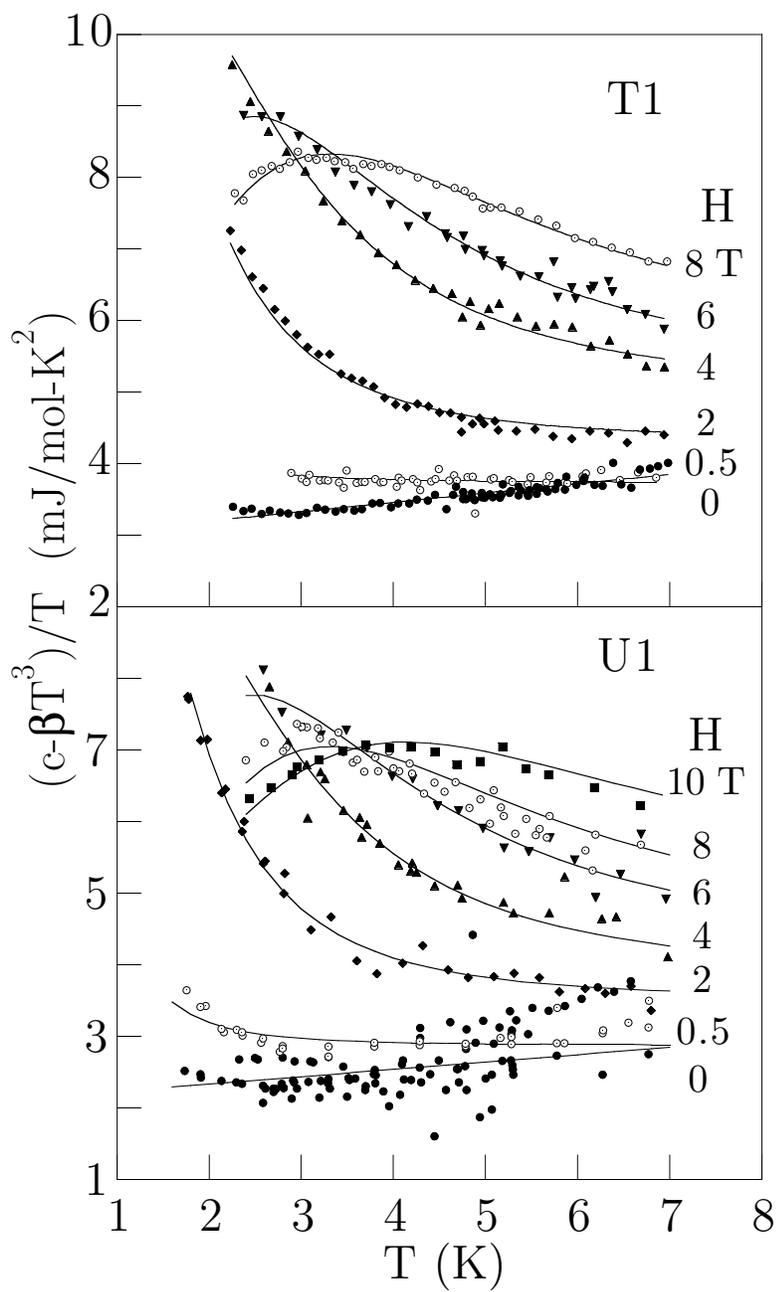
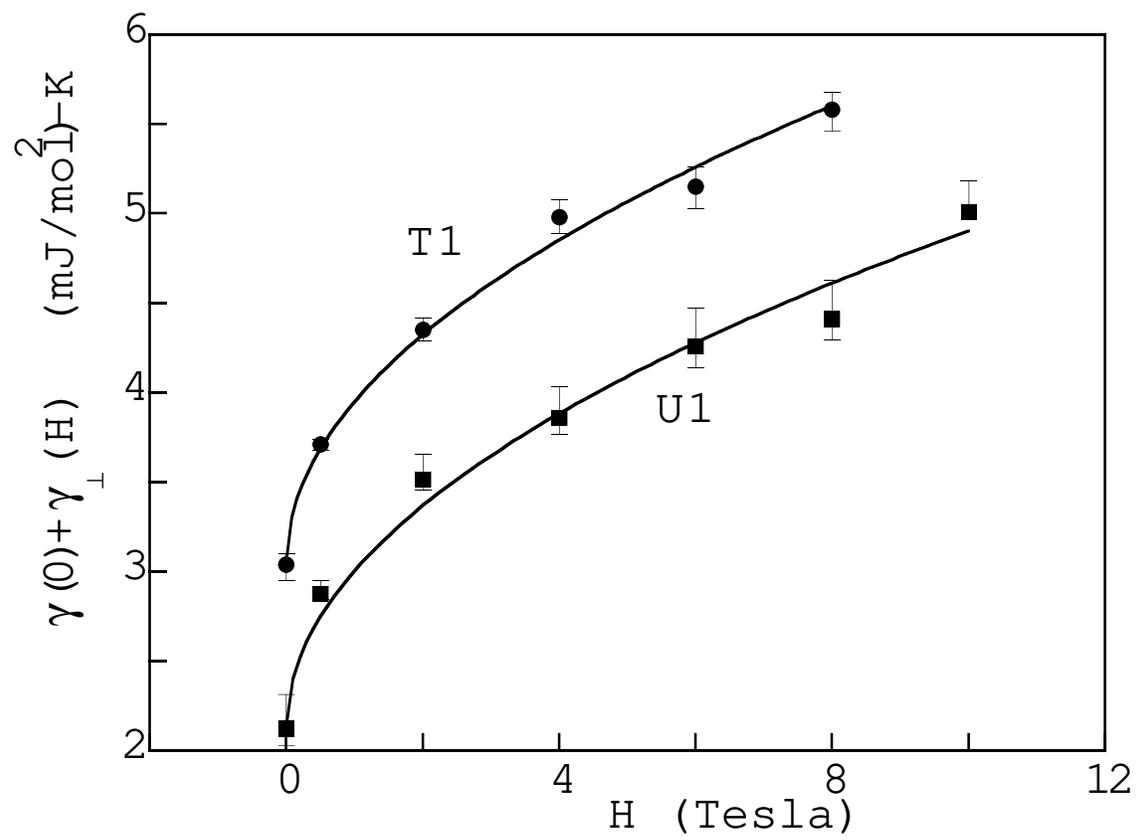
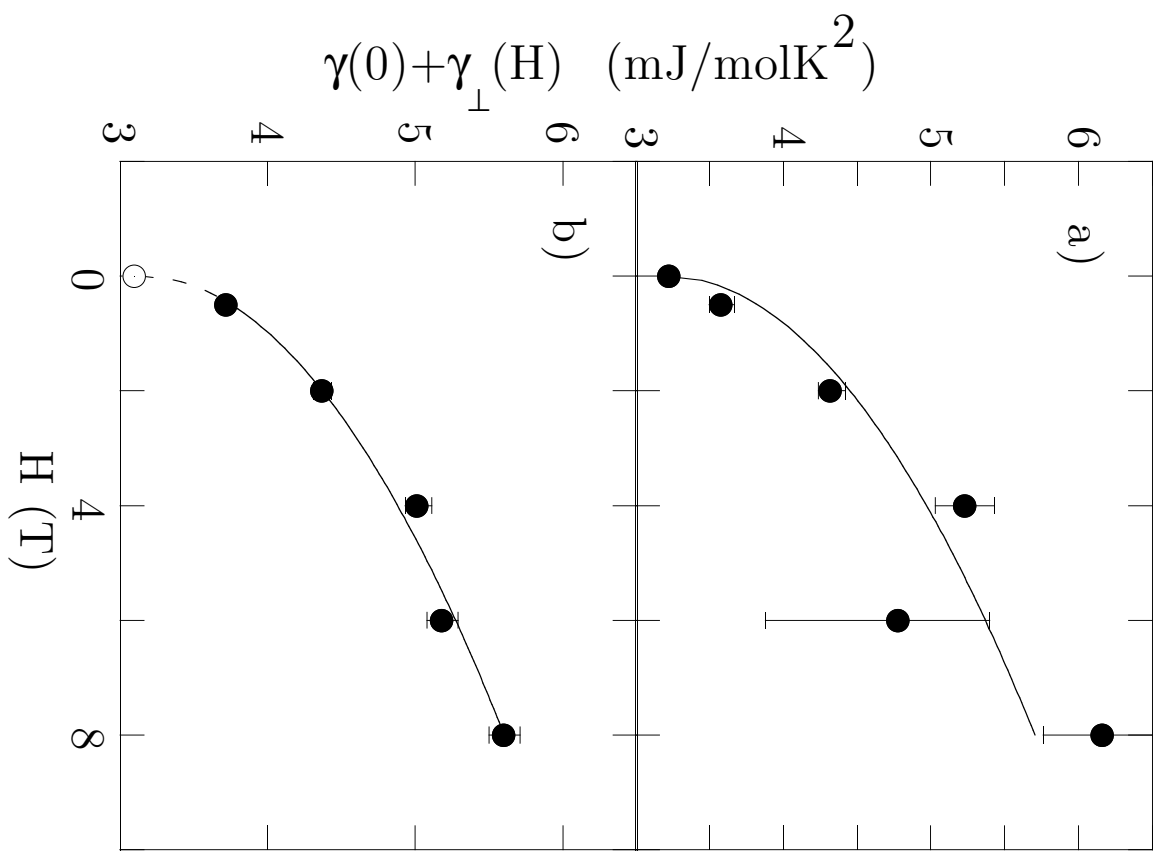
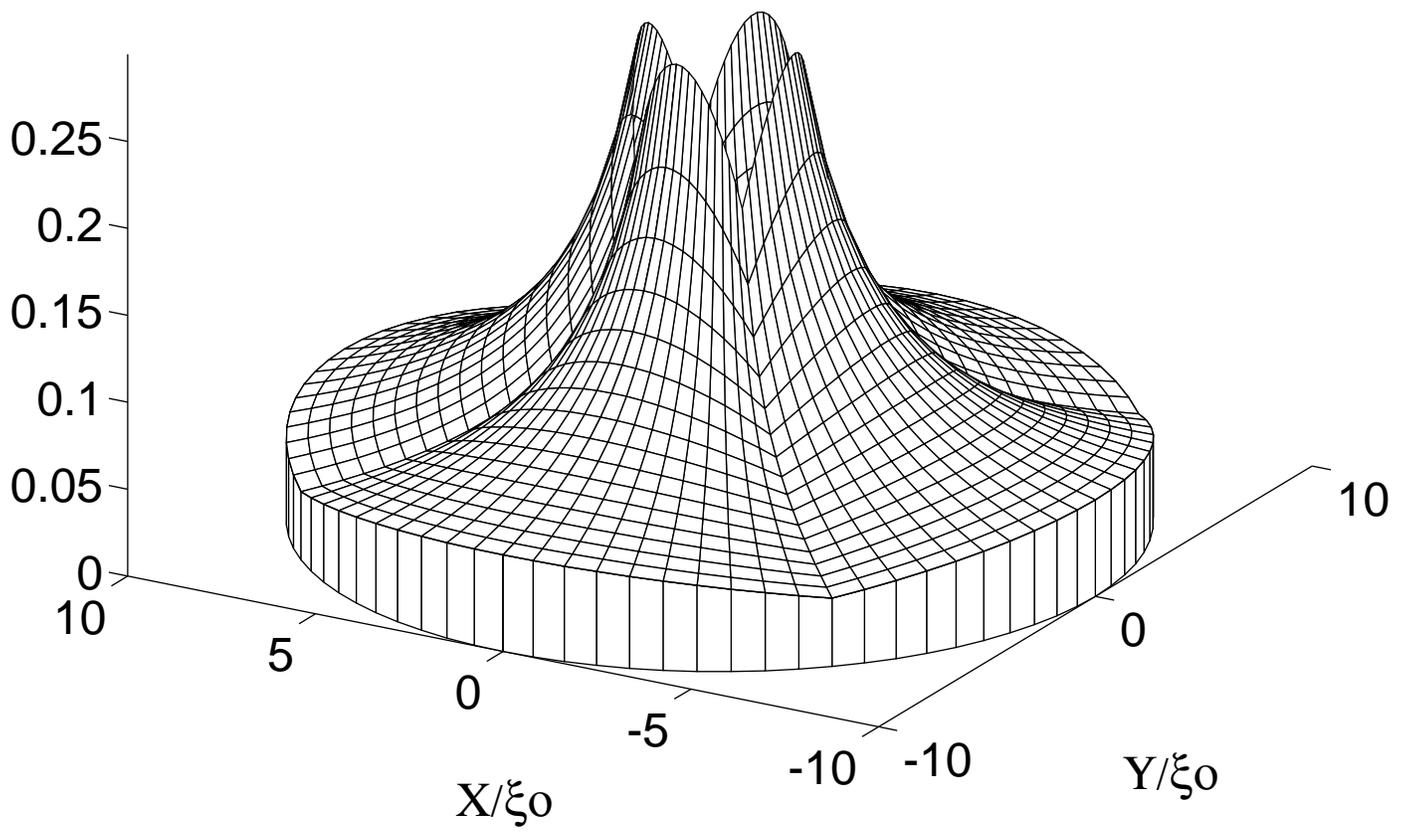
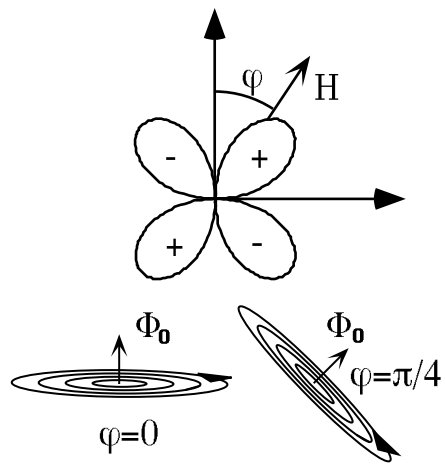
\end{document}